# Plan S. Pardon impossible to execute

Serhii Nazarovets[a], Alexey Skalaban[b]

[a] *Deputy Director for Research at the State Scientific and Technical Library of Ukraine, Kyiv, Ukraine*
[b] *Project Manager Ejournal.by, Minsk, Belarus*

The Plan S initiative is expected to radically change the market of scholarly periodicals, resulting in the abandoning of the subscription model in favour of the open access model. This transition poses new challenges, as well as sets new tasks for researchers, managers, librarians and scientific publishers, particularly in countries with transforming economies. The subscription and publication policies in such countries are still in the making; however, they are to be promptly revised and adjusted in accordance with Plan S.

In this study, we set out to estimate approximate costs and future prospects associated with the transition of Belarus and Ukraine towards open access to scientific information on the example of publications by researchers from these two countries in Elsevier journals. To this end, we accessed the Scopus database and selected all papers affiliated with Belarus and Ukraine published by Elsevier journals in 2018. Subsequently, we established which of these articles indicated a corresponding author from Ukraine or Belarus, as well as collected journal titles, correspondence addresses, access types and the APC fee for each article. In addition, funding sources were established for each open access article. This allowed us to determine the total amount of money that Ukraine and Belarus would have needed to pay so that all the articles of their researchers, which were published in Elsevier journals in 2018, were available in an open-access form.

The obtained results show that Belarus and Ukraine are currently under-investing in the support of scientific publications and subscription to scientific resources, which is necessary for making a transition to the model of open access to scientific literature. There is an urgent need for the authorities in the Belarusian and Ukrainian systems of science to develop national documents regarding open access to scholarly literature, on the basis of which negotiations with leading scientific publishers can be conducted with the purpose of singing transformative agreements. In addition, the current state and future prospects of national open access journals and repositories should be analysed in detail.

The proposed method for calculating the expenses of publishing papers written by researchers from a given country in open access can be used in the preparation of transformative agreements with developing countries.

**Сергей Назаровец** — заместитель директора по научной работе Государственной научно-технической библиотеки Украины, Киев, Украина
**Алексей Скалабан** — руководитель проекта Ejournal.by, Минск, Беларусь

## Plan S. Принять нельзя отказать.

*Реферат:* В недалеком будущем инициатива План S, возможно, в корне изменит рынок научной периодики и мир откажется от подписной бизнес-модели в пользу бизнес-модели открытого доступа. Такой переход ставит перед исследователями, управленцами, библиотекарями и научными издателями из развивающихся стран новые вызовы и задачи, поскольку подписные и публикационные политики этих стран еще только находятся в процессе становления, но уже должны быть оперативно переосмыслены и изменены в соответствии с веяниями Плана S. В этом исследовании на примере публикаций ученых Беларуси и Украины в журналах издательства Elsevier мы попытались оценить приблизительную стоимость и перспективы перехода этих двух стран на открытую модель научной коммуникации. Мы пришли к выводу, что сейчас Беларусь и Украина вкладывают недостаточное количество денег в науку для осуществления перехода на открытую модель. Предложенная нами методика подсчета суммы за публикацию в открытом доступе статей ученых страны может быть использована в процессе подготовки трансформативных соглашений для развивающихся стран.

*Ключевые слова:* Plan S; открытый доступ; Elsevier; научная коммуникация; трансформативные соглашения, Read & Publish

**Введение**

Участники COAlition S, международного консорциума национальных организаций, финансирующих научные исследования при поддержке Европейской Комиссии и Европейского исследовательского совета, в конце 2018 г. разработали инициативу План S (Plan S). Она предусматривает, что после 2021 г. все результаты научных исследований, финансируемых за счет государственных грантов, которые предоставляются Национальным и Европейским исследовательскими советами и органами финансирования, должны публиковаться исключительно в открытых журналах или на соответствующих открытых платформах [1]. Участники COAlition S убеждены, что глобальная научная система может работать эффективно только тогда, когда все результаты исследований полностью доступны научному сообществу. Подписная же модель доступа к научным публикациям скрывает от общества результаты важных научных исследований, тем самым замедляя научный прогресс, и, следовательно, научному сообществу нужно отказаться от финансовой поддержки такой модели.

Члены COAlition S изложили основные идеи Плана S в 10 принципах [2] (*русский перевод принципов был опубликован в газете «Троицкий вариант — Наука»* [3]). На наш взгляд, наиболее важные принципы заключаются в том, что авторы сохраняют права на свои научные работы, которые должны публиковаться в высококачественных открытых журналах и платформах, а исследовательские фонды будут контролировать соответствие этим принципам и накладывать санкции при их нарушении. Иными словами, это был запрет ученым, которые претендуют на получение финансирования от сторонников инициативы COAlition S, публиковаться в журналах, которые не соответствуют принципам Плана S. Так, в первой редакции принципов гибридная модель публикации считалась несоответствующей Плану S, только если журнал не является частью трансформативной сделки по переходу издания на модель золотого открытого доступа.

Этот и некоторые другие принципы Плана были раскритикованы научным сообществом, в том числе и из-за нарушения академических свобод [4–7], и впоследствии в первоначальный механизм реализации Плана S были внесены существенные изменения [8]. Хотя члены COAlition S не будут компенсировать APC в гибридных журналах, исследователям позволено использовать для этого средства из других фондов, и это не будет противоречить требованиям Плана S. Важнейшим достижением в пересмотре руководящих принципов Плана S стала поддержка Open Access 2020 Initiative (OA2020), которая стимулирует научные журналы постепенно отказаться от подписной модели и перейти на модель с открытым доступом [9].

Сторонники Open Access 2020 Initiative обязуются систематически преобразовывать ресурсы, которые в настоящее время расходуются на журнальную подписку, на фонды поддержки стабильных моделей публикации с открытым доступом. Финансовый анализ показывает, что тех денег, которые научные издатели сейчас получают за подписку, вполне достаточно, чтобы покрыть расходы на их переход к публикации с открытым доступом [10]. План S поощряет заключения между учреждениями, консорциумами и издателями соглашений Read & Publish, которые позволяют исследователям в этих учреждениях читать платный подписной контент и публиковать документы в журналах открытого доступа за единую плату. Считается, что такой новый тип соглашений является трансформативным этапом на пути к полному переходу издателя на открытую бизнес-модель.

Хотя основная идея движения Открытого доступа — сделать результаты научных исследований доступными для всех, предложенная в Плане S модель открытого доступа попросту перекладывает финансовые расходы с читателей на авторов научных публикаций, что таит в себе определенные угрозы для ученых из бедных и развивающихся стран. Благодаря реализации плана ученые из развивающихся стран получат более полный и легальный доступ к научным публикациям, но что если при этом они не смогут себе позволить публикацию в открытых журналах, которые соответствуют требованиям Плана S? А что если успешная реализация Плана S переведет этих ученых в ранг исключительно потребителей научного контента? Безусловно, такое развитие событий будет иметь катастрофические последствия для высокотехнологичных и наукоемких отраслей экономики этих стран.

Цель этой работы — на основе данных о публикационной активности ученых за 2018 год посчитать примерную стоимость поддержки Плана S для Беларуси и Украины, оценить выгоды и недостатки перехода этих стран на открытую модель научной коммуникации, которая бы соответствовала требованиям Плана S.

В этом исследовании рассматриваются следующие вопросы:

1. Сколько стоит открыть все научные публикации Беларуси и Украины за 2018 год (на примере журналов издательства Elsevier)?
2. Какой примерно будет плата за публикацию статьи (Article Processing Charge, APC) в ближайшие годы при условии, что План S будет действовать для всех и все журналы будут открыты?

Ответы на эти вопросы позволят нам сделать вывод, достаточно ли сейчас денег в системе науки и образования исследуемых стран для того, чтобы перейти на открытый доступ в 2021 году — год, в котором фонды, которые присоединились к Плану S, планируют требовать от грантополучателей публиковать статьи исключительно в журналах и на платформах, которые гарантируют немедленный открытый доступ к научному контенту.

**Методология**

Исследование проводилось в несколько этапов. Поскольку подсчет стоимости APC для всех статей Беларуси и Украины, которые были опубликованы в разных журналах разных издателей, — процесс очень длительный и трудоемкий, мы решили ограничиться анализом журналов только одного издательства. Поэтому на первом этапе исследования мы определили, сколько статей (тип документа "Article" или "Review") и в каких издательствах опубликовали ученые Беларуси и Украины в 2018 году.

Для этого в базе данных Scopus были отобраны все статьи за 2018 год, которые были аффилированы с Беларусью и Украиной. Поиск проводился 15 июля 2019 г. Поисковый запрос для статей, аффилированных с Беларусью, выглядел как: AFFILCOUNTRY (Belarus) AND (LIMIT-TO (PUBYEAR, 2018)) AND (LIMIT-TO (DOCTYPE, "ar") OR LIMIT-TO (DOCTYPE, "re")). Аналогично выглядел поисковый запрос для статей, аффилированных с Украиной: AFFILCOUNTRY (Ukraine) AND (LIMIT-TO (PUBYEAR, 2018)) AND (LIMIT-TO (DOCTYPE, "ar") OR LIMIT-TO (DOCTYPE, "re")). Списки отобранных документов были экспортированы в CSV-файл, и затем с помощью префиксов DOI было определено название издателя для каждой отобранной публикации.

Анализ показал, что в 2018 году наибольшее количество статей как учеными Беларуси, так и учеными Украины было опубликовано в журналах издательства Springer Nature. Но, поскольку бо́льшая часть журналов, в которых вышли эти статьи, — это переводные версии журналов (т.е. фактически статья подавалась автором в журнал другого издателя), для нашей работы мы решили исследовать журналы издательства Elsevier, которые оказались также весьма популярными каналами распространения научных результатов среди ученых обеих стран (диагр. 1 и 2).

**Диаграмма 1. Распределение публикаций ученых Беларуси в 2018 г. по издательствам (тип документа "Article" или "Review")**

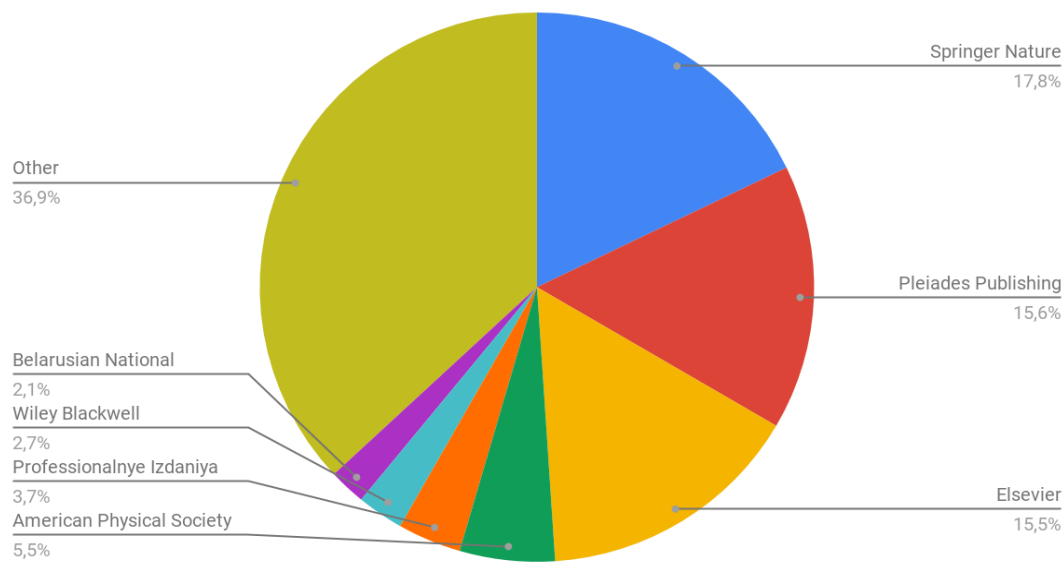

**Диаграмма 2. Распределение публикаций ученых Украины в 2018 г. по издательствам (тип документа "Article" или "Review")**

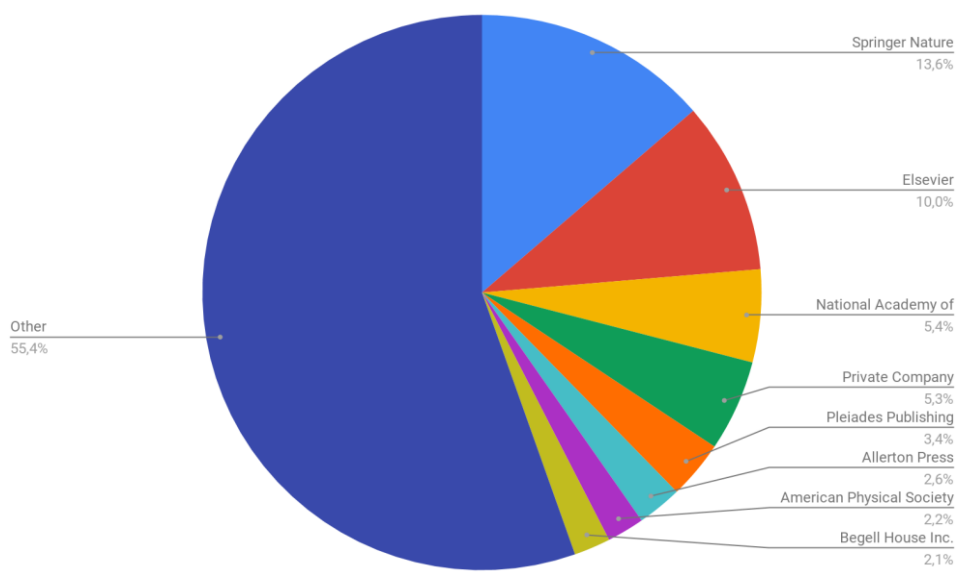

На следующем этапе исследования мы установили, в каких статьях Elsevier был указан автор для корреспонденции (Correspondence Author) из Украины или Беларуси, поскольку именно он отвечает за вопросы, связанные с платой за публикацию статьи (Article processing charge, APC). Для этого с помощью Scopus мы получили файл, который содержал названия публикаций ученых Украины и Беларуси за 2018 год (тип документа "Article" или "Review"), названия журналов, адрес для корреспонденции и тип доступа к каждой статье. Дополнительно для каждой открытой статьи были установлены источники финансирования. С помощью Open Access Price List издательства Elsevier (https://www.elsevier.com/__data/promis_misc/j.custom97.pdf) была установлена стоимость APC для каждой статьи, где автором для корреспонденции был указан автор из Украины или Беларуси. Это позволило нам определить общую сумму, которую должна была бы заплатить Украина и Беларусь для того, чтобы все статьи их ученых в 2018 г. в журналах Elsevier были опубликованы в открытом доступе.

Для оценки перспектив перехода Украины и Беларуси полностью на модель открытого доступа в 2024 году мы попытались:
а) определить, сколько эти страны тратят на подписку доступа к научным журналам в 2019 году;
б) вероятную стоимость сделок типа Read & Publish для Украины и Беларуси, ориентируясь на аналогичные соглашения в других странах. Точные суммы таких потенциальных сделок для Беларуси и Украины неизвестны, и, скорее всего, такая информация является коммерческой тайной.

**Результаты**

*Ориентировочная стоимость открытия всех «эльзевировских» публикаций ученых Беларуси*

Результаты анализа показывают, что в 2018 году ученые Беларуси опубликовали 1912 работ в научных журналах, которые представлены в базе Scopus, из них 605 публикаций представлены в открытом доступе.

Выявлено 297 публикаций с аффилиациями «Беларусь» в журналах издательства Elsevier, из них — 60 публикаций в открытом доступе. В основном, эти открытые статьи являются результатом деятельности больших международных ЦЕРНовских коллабораций: 24 — CMS Collaboration и 20 — ATLAS Collaboration.

В 117 статьях для корреспонденции указан автор из Беларуси. Только 5 из 117 статей были опубликованы в открытом доступе. Обратим внимание на источники оплаты для этих пяти статей. Одна статья опубликована в *Journal of Saudi Chemical Society*. Данное издание относится к категории журналов так называемого «платинового» открытого доступа (Platinum OA), и статьи в нем публикуются за счет King Saud University. Одна статья опубликована в рамках российского проекта 5-100 и оплачена за счет средств Российского университета дружбы народов. За публикацию еще одной статьи, по информации от автора, APC был оплачен за счет соавторов из Турции, и две статьи в *Physics Letters, Section B: Nuclear, Elementary Particle and High-Energy Physics* были оплачены в рамках проекта в области физики высоких энергий SCOAP3 (Sponsoring Consortium for Open Access Publishing in Particle Physics). Кроме этого, еще две статьи опубликованы в журналах, которые не поддерживают открытый доступ.

Анализ стоимости APC для каждой статьи, где автором для корреспонденции был указан автор из Беларуси, позволил установить, что публикация в открытом доступе указанных 110 статей белорусских ученых стоила бы 300 500 долларов США. Средняя стоимость 1 статьи в открытом доступе составляет 2731,8 доллара США.

*Ориентировочная стоимость открытия всех «эльзевировских» публикаций ученых Украины*

Согласно данным базы Scopus, ученые Украины в 2018 году опубликовали 9964 работы в научных журналах, из них — 3200 публикаций представлены в открытом доступе.

Выявлено 998 публикаций, которые аффилированы с Украиной, в журналах издательства Elsevier, из них 116 публикаций в открытом доступе. Открытый доступ для 95 статей ученых Украины был оплачен организациями, которые финансировали исследования, и при этом 61 статья опубликована в журналах «платинового» открытого доступа, в которых публикация в открытом доступе происходит без платы автором за публикацию. Также в одном журнале цена за публикацию в открытом доступе была не указана.

В 473 статьях, которые были опубликованы в журналах Elsevier, был указан автор для корреспонденции из Украины, и только 29 из этих статей были опубликованы в открытом доступе: 16 публикаций в журналах «платинового» открытого доступа, в 9 случаях APC платили сами соавторы, открытый доступ к 4 публикациям оплатили организации, которые финансировали исследования (во всех случаях это было либо совместное, либо исключительно иностранное финансирование). Также 7 статей были опубликованы авторами из Украины в журналах, которые не поддерживают открытый доступ.

После проведенного анализа стоимости APC для каждой статьи, где автором для корреспонденции был указан автор из Украины, мы установили, что публикация 437 статей украинских авторов (без 29 открытых статей в тех 7 в журналах, что не поддерживают ОА) стоила бы 1 147 332 долларов США. Средняя стоимость 1 статьи в гибридном открытом доступе — 2625,5 доллара США.

**Выводы и обсуждения**

Для оценки возможности перехода науки Беларуси и Украины полностью на модель открытого доступа мы дополнительно проанализировали информацию, полученную из открытых источников, о стоимости подписных сделок Elsevier в разных странах, а также попытались объединить различные сведения о текущих государственных расходах Беларуси и Украины на подписку к научным ресурсам, поскольку точный бюджет этих стран на удовлетворение информационных потребностей неизвестен.

В 2019 году 42 университета Беларуси впервые получили полнотекстовый доступ к научным публикациям Elsevier на платформе ScienceDirect. Спонсором этого проекта выступил белорусский бизнесмен Леонид Лознер, цена этой сделки не разглашается [11]. В Украине отсутствует централизованная подписка на полнотекстовые научные ресурсы, в том числе и на платформе ScienceDirect. За средства государственного бюджета оплачивается только доступ к реферативным базам Scopus и Web of Science для университетов и научных учреждений Украины, приблизительная цена этой подписки — 1 690 000 долларов США [12]. Подписка же доступа к полнотекстовым электронным научным ресурсам, которая осуществляется в университетах и научных учреждениях Украины из других источников финансирования, как правило, довольно стохастическая или же просто отсутствует.

Известно, что до недавнего времени Калифорнийский университет платил Elsevier более 10 000 000 долларов США в год за доступ к журналам издательства [13], а Национальный консорциум университетов и научных учреждений Норвегии оформил с Elsevier сделку Read & Publish на два года, которая по предварительным оценкам стоила 10 100 000 долларов США [14]. Полученные результаты свидетельствуют о том, что на государственном уровне Беларусь и Украина пока тратят на всю подписку доступа к научным ресурсам значительно меньше, чем тратит Норвегия на сделку только с одним издательством Elsevier.

В то же время, нам удалось установить, что публикация в открытом доступе всех статей журналов Elsevier в 2018 году, где автором для корреспонденции был указан автор из исследованных стран, была бы для Беларуси и Украины значительно более дешевым вариантом в сравнении со стоимостью упомянутых мировых сделок (300 500 и 1 147 332 долларов США соответственно). Собственно, такая разница в стоимости не удивляет, поскольку в трансформативные соглашения часто и закладывается компенсация издателям издержек перехода на полностью открытую бизнес-модель. Таким образом, университеты и научные учреждения инвестируют в свое будущее. Соответственно, здесь нужно ставить вопрос, какая стоимость подобных трансформативных соглашений будет справедливой для развивающихся стран, и на наш взгляд, предложенная нами методика подсчета суммы за публикацию в открытом доступе всех статей страны за определенный год может служить хорошей отправной точкой в подобных переговорах. Нужно еще упомянуть, что во многих журналах существуют программы поддержки авторов из развивающихся стран (Waiver Policy), благодаря которым эти авторы освобождаются от уплаты APC, поэтому средняя стоимость статьи в гибридном открытом доступе для исследованных стран может быть еще меньше.

Но наш анализ также показал, что Беларусь и Украина тратят на подписку к научным ресурсам значительно меньше даже этой суммы. А также нужно помнить, что статьи в журналах Elsevier составляют приблизительно только десятую часть от всей публикационной активности ученых обеих стран. При этом результаты предыдущих исследований свидетельствуют о том, что ученые Беларуси и Украины активно удовлетворяют свои информационные потребности с помощью нелегальных ресурсов, таких, например, как Sci-Hub [15, 16]. Следовательно, в этих странах ученые нуждаются в доступе к коммерческим научным ресурсам, но этот вопрос должным образом не решен на государственном уровне как Украины, так и Беларуси. Реализация Плана S изменит мир научных коммуникаций, изменятся расходы за доступ к научным статьям и расходы за публикацию научных статей, поэтому роль ресурсов, подобных Sci-Hub, в удовлетворении потребностей ученых существенно девальвирует.

Таким образом, новая система научной коммуникации потребует от чиновников Беларуси и Украины значительно более серьезных глобальных решений, нежели просто нежелание замечать использования учеными страны нелегальных ресурсов. В этой работе мы рассматриваем только статьи, но качественное обеспечение информационных потребностей ученых нуждается также и в доступе к монографиям (особенно в социальных и гуманитарных науках). Потому вопрос о том, как повлияет внедрение Плана S на рынок научных монографий, также является очень важным направлением дальнейших исследований.

Функционерам научных систем обеих стран нужно срочно пересмотреть состояние и перспективы развития собственных национальных открытых журналов, проанализировать их влиятельность, экономическую целесообразность их поддержки, соответствие требованиям Плана S. Стоит также отметить, что сейчас существенно изменилось видение участников COAlition S касательно роли институциональных репозиториев, которые теперь рассматриваются как важные открытые инструменты для распространения научной информации, а не просто как ресурсы для длительного хранения научных документов, как это было в первой редакции Плана S. Количество институциональных репозиториев довольно значительное в Беларуси и Украине, и нужно переосмыслить политику их наполнения и функционирования в условиях развертывания Плана S.

За последние годы многие страны, международные фонды и организации разработали политику и мандаты открытого доступа [17]. Чтобы претворить в жизнь политические стратегии, консорциумы по всему миру начали заключать трансформативные соглашения. Необходимо отметить, что отсутствие государственных документов в области открытого доступа препятствует проведению переговоров с ведущими издателями научных журналов о заключении трансформативных соглашений, поскольку, по опыту других стран, необходимым условием согласия издателей на проведение таких переговоров является существование государственных документов, которые регламентируют перевод в открытый доступ произведений, выполненных при финансовой поддержке государства. Необходимо разработать и принять политику открытого доступа в Украине и Беларуси на национальном уровне, на уровне Академий наук и ведущих государственных научных фондов.

Украина и Беларусь стремятся наращивать свой научно-технический потенциал, что неизбежно должно быть связано с ростом публикационной активности ученых, поэтому для эффективной реализации научной политики управленцам, которые занимаются проблемами развития науки в этих странах, нужно определиться со своим отношением к Plan S. Представления о том, что поддержка производства качественных научных публикаций в современном цифровом мире не требует никаких финансовых затрат или же имеет очень низкую стоимость, — не только ошибочны, но и крайне опасны, поскольку ученым из развивающихся стран может и не найтись места в этом странном новом мире открытого доступа. Поэтому выводы касательно поддержки страной Плана S должны базироваться не просто на уровне аморфных идей открытости, а на реальных расчетах.

**Список литературы**

1. Schiltz M. Why Plan S // https://www.coalition-s.org/why-plan-s/.
2. Principles and Implementation // Website "Science Europe" // https://www.coalition-s.org/principles-and-implementation/.
3. Московкин В. 10 принципов Плана S Евросоюза // Троицкий вариант — Наука. 2018. № 23 (267). С. 5. https://trv-science.ru/2018/11/20/10-principov-plana-s-eu/.
4. Haug C.J. No free lunch. What price Plan S for scientific publishing? // The New England Journal of Medicine. 2019. Vol. 380. No 12. P. 1181–1185. https://doi.org/10.1056/NEJMms1900864.
5. Brainard J. Scientific societies worry Plan S will make them shutter journals, slash services // Science. 2019. Vol. 365. No 6451. https://doi.org/10.1126/science.aaw7718.
6. Spedding M., Barrett J., Morgan E.T., Vore M., Geraghty D., Kirkpatrick C. Plan S: A therat to quality of science? // Science. 2019. Vol. 363. No 6426. P. 462. https://doi.org/10.1126/science.aaw1424.
7. Guzik T.J., Ahluwalia A. Plan S: in Service or Disservice to Society? // European Heart Journal. 2019. Vol. 40. N 12. P. 949–952. https://doi.org/10.1093/eurheartj/ehz065
8. Dal Ré R. Plan S: Funders are committed to open access to scientific publication // European Journal of Clinical Investigation. 2019. Vol. 49. N6. e13100. https://doi.org/10.1111/eci.13100.
9. OA2020 and cOAlition S Launch Joint Statement // Website "Science Europe" // https://www.coalition-s.org/oa2020-and-coalition-s/.
10. Schimmer R., Geschuhn K.K., Vogler A. Disrupting the subscription journals' business model for the necessary large-scale transformation to open access // Max Planck Society. 2015. https://doi.org/10.17617/1.3.
11. Белорусские ученые получили доступ к мировой базе лучших научных публикаций // Веб-портал "Беларусь сегодня" // https://www.sb.by/articles/belorusskie-uchenye-poluchili-dostup-k-mezhdunarodnym-instrumentam.html
12. Лілія Гриневич: Кожен державний виш чи наукова установа зможе підключитися до Scopus та Web of Science за кошти бюджету // Веб-сайт Міністерства освіти і науки України // https://mon.gov.ua/ua/news/liliya-grinevich-kozhen-derzhavnij-vish-chi-naukova-ustanova-zmozhe-pidklyuchitisya-do-scopus-ta-web-science-za-koshti-byudzhetu-dostup-zabezpechat-z-1-chervnya
13. UC Drops Elsevier // Website "Inside Higher Ed's News" // https://www.insidehighered.com/news/2019/03/01/university-california-cancels-deal-elsevier-after-months-negotiations.
14. An Elsevier Pivot to Open Access // Website "Inside Higher Ed's News" // https://www.insidehighered.com/news/2019/04/24/elsevier-agrees-first-read-and-publish-deal.
15. Nazarovets S. Black Open Access in Ukraine: Analysis of Downloading Sci-Hub Publications by Ukrainian Internet Users // Nauka innov. 2018. Vol. 14. No 2. P. 19–26. https://doi.org/10.15407/scin14.02.019.
16. Shvartsman M.E., Lebedev V.V., Skalaban A.V. Sci-Hub as a Mirror of Research and Educational Institutions' Acquisition of E-Resources // Integration of Education. 2017. Vol. 21. No 3. P. 522–534. https://doi.org/10.15507/1991-9468.088.021.201703.522-534.
17. WHO Policy on Open Access // Website "World Health Organization" // https://www.who.int/publishing/openaccess/en/.